%% file: main.tex
\documentclass[journal]{IEEEtran}
\usepackage{amsmath,amssymb,amsfonts,mathrsfs,mathtools,bm, bbm, dsfont,mathrsfs,amsthm,blkarray}
\usepackage[dvipdfm]{graphicx}
\usepackage[dvipsnames]{xcolor}

\usepackage{epsfig,epsf,psfrag,latexsym}

\usepackage{booktabs}
\usepackage{multirow,multicol}

\usepackage[breaklinks,colorlinks, linkcolor=MidnightBlue, anchorcolor=MidnightBlue, citecolor=MidnightBlue, urlcolor=MidnightBlue]{hyperref}
\usepackage[hyphenbreaks]{breakurl}
\usepackage{cite}
\usepackage{xurl}

 \def\old#1{}    % Please don't remove this... This command includes the text to be deleted.

\input LTmacros

\input{header.tex}
\title{Decentralized Equitable Energy Access \\ in Energy Communities}
\author{Siying Li,\IEEEmembership{} Timothy D. Mount, Lang Tong\IEEEmembership{}
\thanks{\scriptsize  This paper is to be published in the proceedings of the 2024 60th Annual Allerton Conference on Communication, Control, and Computing.}
\thanks{\scriptsize Siying Li and Lang Tong (\url{{sl2843 ,lt35}@cornell.edu}) are with the School of Electrical and Computer Engineering, Cornell University, Ithaca, NY 14853, USA. Timothy Douglas Mount (\url{tdm2@cornell.edu}) is with the Dyson School of Applied Economics and Management, Cornell University, Ithaca, NY 14853, USA. This work was supported in part by the National Science Foundation under Grants 2218110 and 2412776.}
}
\begin{document}
\maketitle

\begin{abstract}
We address the issue of equitable energy access within an energy community consisting of members with diverse socioeconomic backgrounds, including varying income levels and differing capacities to access distributed energy resources such as solar power and storage systems. While optimal energy consumption scheduling is well-studied, integrating equity into decentralized real-time energy access remains under-explored. This paper formulates Equity-regarding Welfare Maximization (EqWM)---a welfare optimization energy scheduling subject to equity constraints. We further develop a decentralized implementation (D-EqWM) as a bi-level optimization, where a non-profit operator designs a community pricing policy aimed at maximizing overall welfare, subject to constraints that ensure equitable access. Community members, in turn, optimize their individual consumption based on these prices. We present the optimal pricing policy along with its key properties.
\end{abstract}


\section{Introduction}\label{sec:Intro}

\input{intro}
% We consider equitable access to energy in an energy community (EC) consisting of members from different socioeconomic backgrounds, such as unequal income levels and unequal abilities to access energy resources such as solar, storage, etc. Although energy consumption scheduling is well studied, the literature on incorporating equity measures in optimal energy access in real-time---the so-called {\it ex post\footnote{Ex ante equitable access deals with equity measures before randomness is realized. Ex post equitable access deals with equity with realized energy access.}} equitable energy access---is lacking.
We define energy community (EC)  as a collection of $N$ community members in the same service area of a distribution utility. Some members are prosumers with private distributed energy resources (DER) such as rooftop solar or behind-the-meter storage while others do not have access to DER. By incorporating community pricing policies that promote equitable energy access, a community member with no possibility of having rooftop installations ({\it e.g.,} an apartment renter) or a low-income household without the financial means of investing in rooftop solar can benefit from DER installations.

We assume that energy access in an EC is managed by a non-profit EC operator, playing the role of an energy aggregator who manages energy consumption and production within the community,  purchasing power from and exporting surplus power to the distribution utility under the regulated net energy metering (NEM) tariff.

This paper focuses on equitable access to DER within a community via community-pricing-based decentralized consumption scheduling. The underlying problem addressed here is also relevant to energy justice which requires the community to adequately serve its most disadvantaged members and some form of fairness in energy access. The EC operator sends pricing signals to all members, based on which every member optimizes its consumption, taking into account the available energy from its private DER and its budget constraint tied to its income level. The goal of the EC operator is to design a community pricing mechanism that influences the consumption of its members to achieve several objectives: (1) {\em Efficiency}---maximizing community social welfare defined as the sum of consumer surplus of all members, (2) {\em Individual rationality}---guaranteeing each member gains no less surplus within the community than being served by the distribution utility so that no member abandons the community, (3) {\em Operator revenue adequacy}---revenue from members matches the payments to the distribution utility, and (4) {\em Equity standard}---ensuring the community meets an equity standard,  measured by the Rawlsian social welfare function. 
% The last objective separates this work from the standard decentralized welfare-maximizing scheduling problems in the literature. The community pricing satisfying the above objectives defines a framework of decentralized {\em Equity-regarding Welfare Maximization (EqWM)}.

\section{Equity Measure for Energy Access}
In their pioneering work, Pigou \cite{Pigou:12book} and Dalton \cite{dalton_measurement_1920} articulated the fundamental principle of wealth transfer in addressing income inequality. The so-called Pigou-Dalton principle states that an income transfer from a wealthy individual to a poorer one reduces income inequality.
Making a connection between the Pigou-Dalton principle with Lorenz curve and using an earlier result of Rothschild and Stiglitz on distribution risks \cite{rothschild_increasing_1970}, Atkinson showed in \cite{atkinson_measurement_1970} that an income distribution $A$ has an equal or higher level of equity than $B$ under the Pigou-Dalton principle if and only if the Lorenz curve of $A$ is nowhere under that of $B$.

Herein, we use the notation $f_A \stackrel{ \mbox{\sf \tiny PD}}{\succcurlyeq} f_B$ to mean that $f_A$ exhibits a higher level of equity than $f_B$, thus making it preferred in the Pigou-Dalton sense. In other words, $f_A \stackrel{\mbox{\sf \tiny PD}}{\succcurlyeq} f_B$ implies that the Lorenz curve associated with $f_A$ lies everywhere on or above that of $f_B$. Note that relation $\stackrel{\mbox{\sf \tiny PD}}{\succcurlyeq}$ establishes only a partial order among distributions.

Incorporating the Lorenz curve in optimization, however, is nontrivial. Alternatively, social welfare functions (SWF) have been considered as measures for equity by Dalton \cite{dalton_measurement_1920}, Atkinson \cite{atkinson_measurement_1970}, and others \cite{dasgupta_notes_1973, shorrocks_ranking_1983}. Given the income (or energy consumption vector) of $N$ households $\dbf=(d_1,\cdots, d_N)$,
a social welfare function $\mbox{SWF}$ is a function of utilities of individual consumption $\dbf$ defined by
\begin{equation}
    W=\mbox{\sf SWF}(\tilde{U}(\dbf)),~~\tilde{U}(\dbf):=(\tilde{U}(d_1),\cdots,\tilde{U}(d_N)),
\end{equation}
where $\tilde{U}$ is a common utility function of consumption.  The specific forms of $W$ and $\tilde{U}$ turn out to be immaterial. Dasgupta, Sen, and Starrett showed in \cite{dasgupta_notes_1973} that,  as long as  $\tilde{U}$ is concave and $W$ increasing, symmetric, and quasi-concave in individual utilities, for every such $W$,  distributions that are partially ordered by the Pigou-Dalton principle can be ranked by the SWF, \ie
\begin{equation}
\dbf^A \stackrel{\mbox{\sf \tiny PD}}{\succcurlyeq} \dbf^B~~\Rightarrow~~
\mbox{\sf SWF}(\tilde{U}(\dbf^A))\ge \mbox{\sf SWF}(\tilde{U}(\dbf^B)).
\end{equation}
That any SWF satisfying the above mild conditions can be used to rank distributions according to the Pigou-Dalton principle makes SWF a convenient choice to impose equity constraints.  In particular, by imposing 
\begin{equation}
    \mbox{\sf SWF}(\tilde{U}(\dbf))\ge \omega
\end{equation}
in an optimization, where $\omega$ is the equity standard representing the minimum acceptable level of equity, the solution $\dbf^*$ is guaranteed to be PD-preferred over any $\dbf$ with smaller SWF whenever $\dbf^*$ and $\dbf$ are comparable, \ie  $\dbf^*$ is reachable via mean-preserving welfare transfer from $\dbf$.

In this paper, we adopt the Rawlsian social welfare function 
\begin{equation}
    \mbox{\sf SWF}(\tilde{U}(\dbf)) = \min\{\tilde{U}(d_1),\cdots, \tilde{U}(d_N)\},
\end{equation}
which uses the worst-off member as a way to measure equity of consumption. Note that using the Rawlsian social welfare function allows us to interpret the results here in the context of energy justice and fairness by taking Rawls' perspective of assuming ``complete ignorance" of social position in evaluating justice and his notion of justice as fairness \cite{Rawls1999}.

\section{Equity-regarding Welfare Maximization}
This section introduces Equity-regarding Welfare Maximization (EqWM) and a decentralized implementation referred to as D-EqWM.
\subsection{EqWM: Centralized Implementation}
In the framework of centralized optimization, the operator schedules all resources for the benefit of the community. The energy consumption $d_i$ and the corresponding payment $p_i$ for each member are determined by optimization \eqref{eq:optcent} and announced. Centralized decision-making ensures that the community's total surplus cannot be further improved. 
\begin{equation}\label{eq:optcent}
    \begin{array}{lll}
    \underset{\{(d_i,p_i)\}}{\rm max}&& \sum_{i=1}^N U_i(d_i)-\Pi^{\scalebox{0.55}{NEM}}(\sum_{i=1}^N(d_i-g_i)) \\
     \mbox{subject to} && \Pi^{\scalebox{0.55}{NEM}}(\sum_{i=1}^N(d_i-g_i))=\sum_{i=1}^N p_i,\\
     && {\rm min}_{i\in \{1,2,...,N\}}\tilde{U}(d_i)\geq \omega,\\
     % &\alpha: &\sum_{i=1}^N \log(U_i(d_i))\geq \omega,\\
     && \forall i \in \{1,2,...,N\},\\
     % && \tilde{U}(d_i)\geq \omega,\\
     && p_i\leq x_i,\\
     && U_i(d_i)-p_i\geq s_i^{\scalebox{0.7}{out}}(g_i).
    \end{array}
\end{equation}
where $U_i$ is the utility function of member $i$ for consuming energy, $g_i$ is member $i$'s renewable generation, and $x_i$ represents its income-based energy budget. $s_i^{\scalebox{0.7}{out}}(g_i)$ is the surplus of member $i$ as a standalone consumer under utility rate, which can be computed explicitly from the following optimization:
\begin{equation}\label{eq:benchmark}
    \begin{array}{lll}
    \underset{d_i}{\rm max}&& U_i(d_i)-\Pi^{\scalebox{0.55}{NEM}}(d_i-g_i) \\
     \mbox{subject to}
     && \Pi^{\scalebox{0.55}{NEM}}(d_i-g_i)\leq x_i.
    \end{array}
\end{equation}
We enforce that the surplus of each member in the EC is no less than it would be as a standalone consumer. $\Pi^{\scalebox{0.55}{NEM}}(z)$ is the payment function for net electricity consumption $z$ under the regulated NEM tariff,
\begin{equation}
    \Pi^{\scalebox{0.55}{NEM}}(z):= \left\{
    \begin{array}{lcl}
    \pi^+z && z\geq 0\\
    \pi^-z && z\leq 0\\    
    \end{array}\right..
\end{equation}
Here, $\pi^+$ is the buy rate when the customer net-consumes. When the customer is a net-producer, it is compensated by the sell rate $\pi^-\leq \pi^+$.

\subsection{D-EqWM: A Decentralized Implementation}
D-EqWM is a decentralized implementation of EqWM with a bi-level optimization.  At the upper level is the community operator who designs {\em pricing policy} $\psi$ that maps available renewable generation $\gbf=(g_i)$ to a set of pricing parameters $\{\thetabf_i\}$ that define community members' payment functions $\Pi_i (d_i,g_i) \mapsto p_i$.  Note that, by the Pigou-Dalton principle, discriminative pricing is necessary to achieve a level of equity.
At the lower level, upon receiving $\thetabf_i$, member $i$ optimizes its consumption subject to its budget constraint $x_i$.

% Decentralized equity-regarding welfare maximization (EqWM) deals with decentralized energy consumption scheduling when measurements of renewable generation $\mathbf{g}=(g_i),~ i\in\{1,2,...,N\}$ are available in real-time scheduling. At the beginning of each scheduling interval, the operator determines the community consumption prices $\thetabf_i$ based on a community pricing policy $\psi=(\psi_i)$ and broadcasts the payment function $\Pi_i (d_i|\thetabf_i,g_i)$ to each member. Given $\Pi_i (d_i|\thetabf_i,g_i)$, each member sets its own consumption following the individual consumption policy $\chi_i$. The problem of decentralized EqWM is to determine the optimal pricing and consumption policies as the optimal solution for bi-level optimization, as follows. To incorporate equity considerations, we impose an energy budget constraint on individual members and an equity constraint on the EC.
\subsubsection{Member consumption policy $\chi_i$} The lower-level optimization \eqref{eq:optcons} is employed to optimally schedule the EC members' energy consumption.
% Without loss of generality, suppose that each member has a single flexible demand scheduled by its consumption policy. Given the realized individual renewable generation $g_i$ and the broadcasted pricing policy represented by a payment function $\Pi_{i}$ from the operator, member $i$ schedules its consumption.

While ensuring that the payment to the operator does not exceed its budget $x_i$, member $i$ maximizes the consumption surplus $S_i(d_i)$\footnote{We denote $S_i(d_i)$ as the simplified form of $S_i(d_i| \thetabf_i,g_i)$.}, defined by the utility $U_i(d_i)$ of consuming $d_i$ minus the payment.
\begin{equation}\label{eq:optcons}
    \begin{array}{lll}
    \chi_i(\thetabf_i,g_i):= &\underset{\substack{d_i}}{\arg \max}& S_i(d_i)\\
    & \mbox{subject to}&   S_i(d_i)=U_i(d_i)- \Pi_i (d_i| \thetabf_i,g_i),\\
    && \Pi_i (d_i| \thetabf_i,g_i)\leq x_i.
    \end{array}
\end{equation}
% where $U_i(\cdot)$ is the utility function of member $i$ for consuming energy, it is assumed to be concave. 
% $d_i$ and $g_i$ denote the energy consumption and realized renewable of member $i$, respectively.
% where $\mathcal{D}_i$ is the set of feasible consumption for member $i$.

\subsubsection{Community pricing policy $\psi$} 
The operator's pricing policy $\psi$ defines the payment function $\Pi_{i} (d_i|\thetabf_i,g_i)$ for each member. Through the upper-level optimization \eqref{eq:optprice}, the EC operator maximizes its members' total surplus while enforcing a minimum acceptable level of energy access equity within the community.
\begin{equation}\label{eq:optprice}
    \begin{array}{lll}
    \psi(\textbf{g}):=&\underset{\substack{(\thetabf_i)}}{\arg \max}&\sum_{i=1}^NU_i(\chi_i(\thetabf_i,g_i))-\Pi^{\scalebox{0.55}{NEM}}(z)\\
    &\mbox{subject to} & z=\sum_{i=1}^N (\chi_i(\thetabf_i,g_i)-g_i),\\
    &&\Pi^{\scalebox{0.55}{NEM}}(z)=\sum_{i=1}^N \Pi_{i}(\chi_i(\thetabf_i,g_i)| \thetabf_i,g_i),\\
    % && \underset{i\in \{1,2,...,N\}}{\rm min}\tilde{U}(\chi_i(\thetabf_i,g_i))\geq \omega,\\
    && {\rm min}_{i\in \{1,2,...,N\}}\tilde{U}(\chi_i(\thetabf_i,g_i))\geq \omega,\\
    && \forall i \in \{1,2,...,N\},\\
    && S_i(\chi_i(\thetabf_i,g_i))\geq s_i^{\scalebox{0.7}{out}}(g_i).
    \end{array}
\end{equation}
% where $\omega$ is the equity standard, representing the minimum acceptable level of equity for the EC. $s_i^{\scalebox{0.7}{out}}(g_i)$ is the surplus of member $i$ as a standalone consumer under utility rate. We enforce that the surplus of each member in the EC is no less than it would be as a standalone consumer. $\Pi^{\scalebox{0.55}{NEM}}(z)$ is the payment function for net electricity consumption $z$ under the regulated NEM tariff,
% \begin{equation}
%     \Pi^{\scalebox{0.55}{NEM}}(z):= \left\{
%     \begin{array}{lcl}
%     \pi^+z && z\geq 0\\
%     \pi^-z && z\leq 0\\    
%     \end{array}\right..
% \end{equation}
% Here, $\pi^+$ is the buy rate when the customer net-consumes. When the customer is a net-producer, it is compensated by the sell rate $\pi^-\leq \pi^+$.

\section{Optimal Pricing Policy and Properties}
We highlight the D-EqWM solution and its properties in this section, leaving detailed theoretical development in \cite{LiMountTong}.

\subsection{Optimal D-EqWM Policy}
The optimal member consumption policy is defined by \eqref{eq:optcons}, and it is easily obtained if the pricing policy is convex. To this end, we will restrict ourselves to the class of affine pricing policies where the payment function $\Pi_i$ to be affine parameterized by $\thetabf_i=(\theta_i,\theta_0)$ with uniform volumetric price $\theta_0$ and discriminative fixed charge $\theta_i$. Specifically, if the consumer has net-consumption $z_i=d_i-g_i$, it pays to the community operator by
\begin{equation}
    \Pi_i(z)=\theta_i + \theta_0 z.
\end{equation}
It is shown in \cite{LiMountTong} that, under mild conditions, such specialization is made without loss of generality.

The optimal community pricing policy is defined by the optimal pricing parameters $\thetabf_i^*=(\theta_i^*,\theta^*_0)$. The optimal volumetric price $\theta_0^*$ and fixed charges ($\theta_i^*$) play separate roles in D-EqWM. Specifically, $\theta_0^*$ maximizes social welfare subject to budget and individual rationality constraints. On the other hand, the discriminative fixed charges implement a version of Pigou-Dalton welfare transfer.  Both prices can be computed explicitly.

\begin{figure}[h]
    \centering
    \includegraphics[scale=0.45]{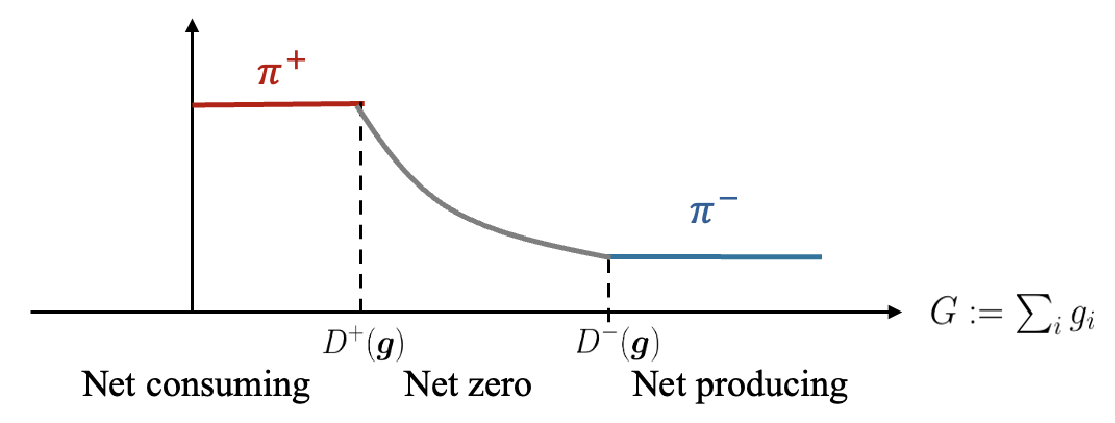}
    \caption{\scriptsize Optimal volumetric price $\theta_0^*$ under D-EqWM.}
    \label{fig:volumetric}
\end{figure}

The optimal volumetric price has a simple structure as shown in Fig. \ref{fig:volumetric}, where $\theta_0^*$ is defined on the total renewable generation $G=\sum_i g_i$ in three regions. When  $G=\sum_i g_i < D^+(\gbf)$,  the community-wide price of consumption/production is the utility's retail (net-consumption) price $\pi^+$, resulting in the community as a whole being a net-consumer.  When  $G> D^-(\gbf)$,  the community-wide price is the utility's net-producing price $\pi^-$, resulting in the community as a whole being a net-producer. In between, the community price is $\pi^0(\gbf)$, resulting in the community as a whole net-zero consumer.

The optimal fixed charges can be solved from a convex optimization where the Rawsial equity constraint is achieved through mean-preserving transfers. It is this process that guarantees improved equity in the Pigou-Dalton sense.

\subsection{Properties of D-EqWM}
We establish several properties of D-EqWM in \cite{LiMountTong}. First, on efficiency,  we show that the optimal D-EqWM solution achieves the social welfare of the centralized EqWM, despite the use of a specialized affine pricing policy.

Second, by definition of the decentralized optimization, D-EqWM also ensures that every member within the community gains surplus no less than achievable being outside the community, making the community having competitive advantage over the utility pricing. 

\begin{figure}[h]
    \centering
    \includegraphics[scale=0.6]{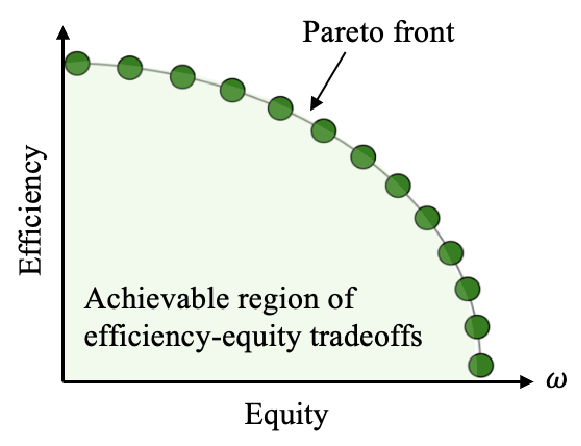}
    \caption{\scriptsize Pareto front of the efficiency-equity tradeoff for D-EqWM.}
    \label{fig:tradeoff}
\end{figure}

Third, we show that the region of achievable efficiency-equity level is concave with every point on the Pareto front achievable by D-EqWM as shown in Fig. \ref{fig:tradeoff}. In particular, as equity measure increases from $\omega_1$ to $\omega_2$, the optimal consumption distribution $\dbf_2^*$ from D-EqWM is Pigou-Dalton preferred over $\dbf_1^*$.

\section{A Numerical Example}
To demonstrate the performance of D-EqWM, we considered an energy community with $N=100$ members. Each member has flexible loads, and 75 of them have rooftop solar installations. Individual utilities are assumed to be increasing quadratic functions. In the regulated NEM tariff, the buy rate $\pi^+=\$0.4{\rm /kWh}$ and the sell rate $\pi^-=\$0.2{\rm /kWh}$. 

In our simulation, scheduling decisions were executed over a 1-hour horizon. The energy budget distribution follows the 2020 annual household site consumption and expenditure data, released by the U.S. Energy Information Administration \cite{energydata}. We randomly generated 100 budget realizations based on this distribution for the EC. Within each budget realization, 100 random renewable generation scenarios were sampled, with the solar generation for each solar-owning household calculated as the sum of the predicted solar output and a Gaussian-distributed error.

We compared the optimal energy consumption of members under three pricing policies: 1) Each household operates as a standalone customer under DSO's NEM tariff; 2) households are within the EC under the NEM pricing; and 3) households act as EC members, and the D-EqWM solution is implemented.

For each pricing policy, we calculated every member's expected energy consumption across sampled renewable generation scenarios for each budget realization. We then took the expectation over individual budgets to obtain the expected individual energy consumption.
\begin{figure}[h]
    \centering
    \includegraphics[scale=0.3]{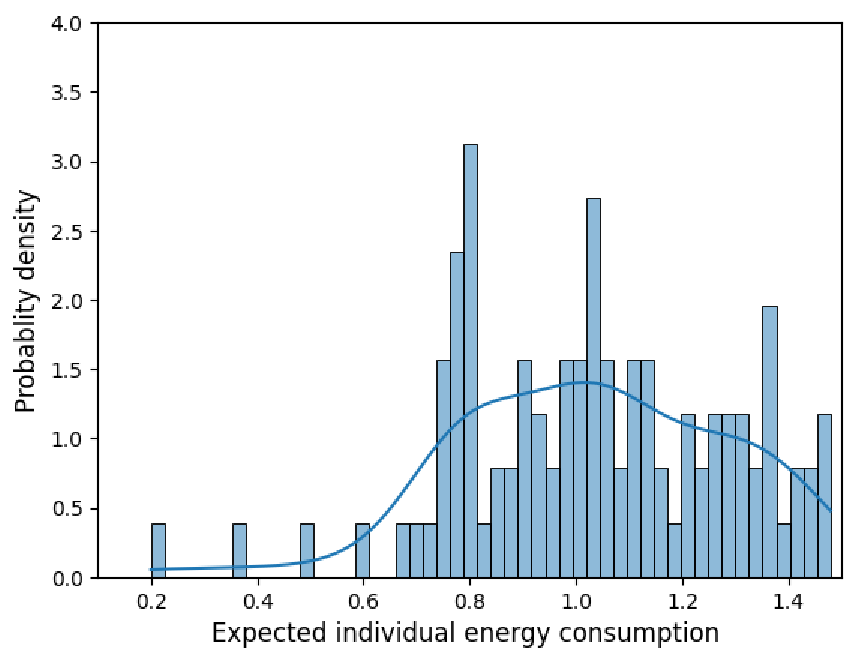}
    \includegraphics[scale=0.3]{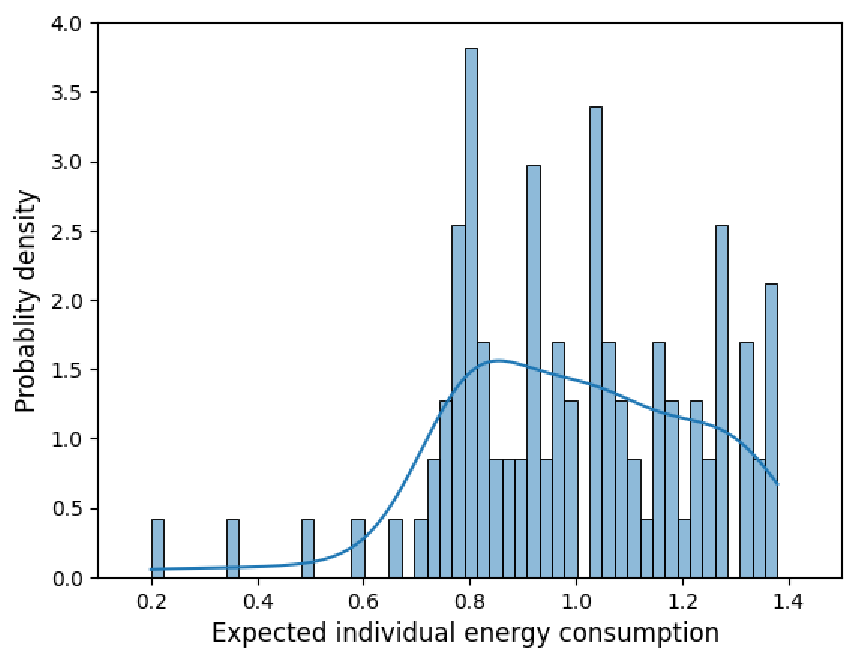}
    \includegraphics[scale=0.3]{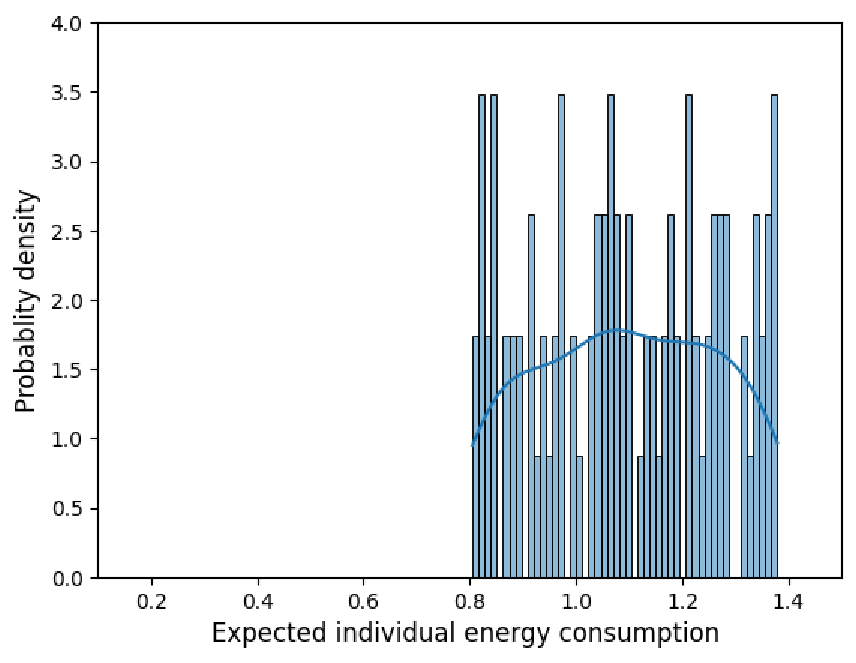}
    \includegraphics[scale=0.28]{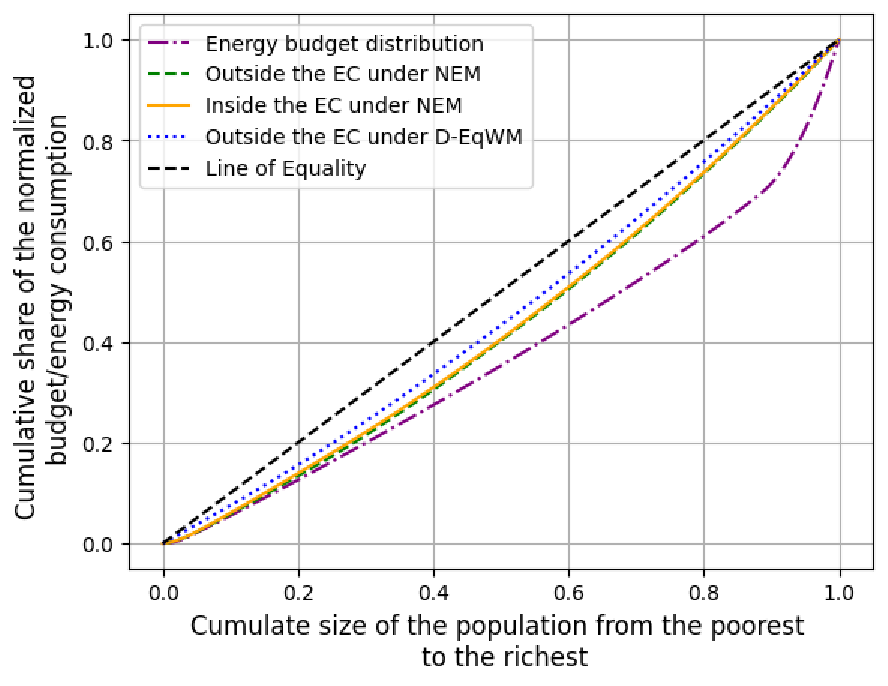}
    \caption{\scriptsize Distributions of expected energy consumption under different pricing policies and their corresponding Lorenz curves. (Top left: expected energy consumption distribution for households outside the EC under NEM tariff; top right: expected energy consumption distribution for households inside the EC under NEM tariff; bottom left: expected energy consumption distribution for households inside the EC under D-EqWM; bottom right: Lorenz curves of energy budget and consumption distributions).}
    \label{fig:equity}
\end{figure}

As shown in Fig. \ref{fig:equity}, without discriminative pricing, the equity of the expected energy consumption distribution does not improve significantly after households join the EC. However, under the proposed D-EqWM framework, the expected energy consumption among all EC members is more concentrated, implying higher energy access equity. This result is supported by the Lorenz curves of these distributions, as the Lorenz curve of the energy consumption distribution under D-EqWM is the closest to the line of equality.
\begin{figure}[h]
    \centering
    \includegraphics[scale=0.45]{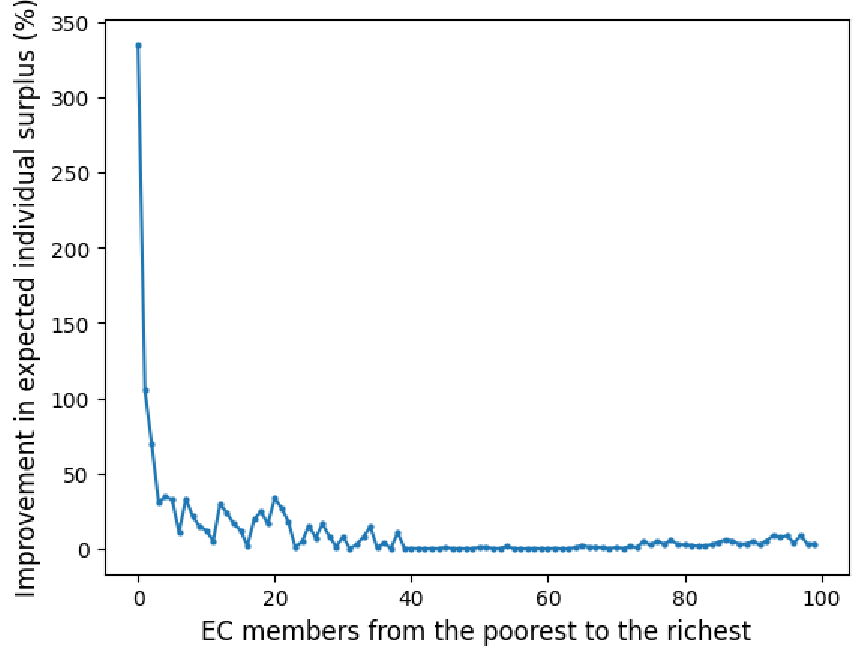}
    \caption{\scriptsize Improvement in expected individual surplus under D-EqWM.}
    \label{fig:surplus}
\end{figure}

Fig. \ref{fig:surplus} illustrates the improvement in expected surplus for each EC member under D-EqWM compared to being standalone outside the EC. It shows that joining the EC under the D-EqWM framework allows each member to gain a surplus that is no less than what could be achieved outside the community, ensuring that no one abandons the community. Notably, members with limited energy budgets experience a more significant improvement in surplus. This is because the Pigou-Dalton welfare transfer is implemented to help them consume more energy.

\section{Conclusions}\label{sec:Conclusion}
We developed equity-regarding welfare maximization (EqWM) and a decentralized implementation (D-EqWM) for scheduling energy consumption in a community characterized by income disparities and unequal access to energy resources. D-EqWM ensures that the community meets a predefined equity standard, as measured by the Rawlsian equity criterion, while guaranteeing that each member achieves a consumption surplus no lower than what they would receive under the utility's NEM tariff. A key insight is that, while the widespread adoption of rooftop solar tends to exacerbate energy inequalities, an energy community can benefit all members—rich and poor alike—by improving equitable access to distributed energy resources. The proposed EqWM also provides a Rawlsian interpretation of energy justice and fairness.

% This paper presents decentralized EqWM, a scheduling framework for energy communities to maximize community social welfare while ensuring a certain level of energy access equity. Through decentralized EqWM, the EC operator sends dynamic prices to members based on renewable generation. Each member then optimizes the individual energy consumption in response to the broadcasted prices. We demonstrate that the equilibrium of this bi-level optimization exists and can be achieved by the proposed discriminative pricing mechanism. In addition to ensuring individual rationality for members and revenue adequacy for the EC operator, the optimal solution of decentralized EqWM achieves the maximum total surplus of centralized optimization and allows for improvement in energy access equity according to the Lorenz curve ordering.

{
\bibliographystyle{IEEEtran}
\bibliography{BIB}
}

% \newpage
% \[

% \]
% \newpage

% \section{Appendix}

\end{document}

%% file: LTmacros.tex
\def\beq{\begin{equation}}
\def\eeq{\end{equation}}
\def\bea{\begin{eqnarray}}
\def\eea{\end{eqnarray}}
\def\ba{\begin{array}}
\def\ea{\end{array}}

\def\bitem{\begin{itemize}}
\def\eitem{\end{itemize}}
\def\ben{\begin{enumerate}}
\def\een{\end{enumerate}}

\def\ie{{\it i.e.,\ \/}}

\def\thetabf{{\mbox{\boldmath$\theta$\unboldmath}}}
\def\dbf{{\bm d}}

\def\gbf{{\bm g}}

%% file: header.tex
\newcommand{\beqa}{\begin{eqnarray}}
\newcommand{\eeqa}{\end{eqnarray}}
\newcommand{\beqan}{\begin{eqnarray*}}
\newcommand{\eeqan}{\end{eqnarray*}}

\newcounter{l1}
\newcounter{l2}
\newcounter{l3}
\newcommand{\bdotlist}{\begin{list}{$\bullet$}{}}
\newcommand{\bboxlist}{\begin{list}{$\Box$}{}}
\newcommand{\bbboxlist}{\begin{list}{\raisebox{.005in}{{\tiny
$\blacksquare$ \ \ }}}{}}
\newcommand{\bdashlist}{\begin{list}{$-$}{} }
\newcommand{\blist}{\begin{list}{}{} }
\newcommand{\barablist}{\begin{list}{\arabic{l1}}{\usecounter{l1}}}
\newcommand{\balphlist}{\begin{list}{(\alph{l2})}{\usecounter{l2}}}
\newcommand{\bAlphlist}{\begin{list}{\Alph{l2}.}{\usecounter{l2}}}
\newcommand{\bdiamlist}{\begin{list}{$\diamond$}{}}
\newcommand{\bromalist}{\begin{list}{(\roman{l3})}{\usecounter{l3}}}